\documentstyle[aps]{revtex}

\begin{document}
\title{Spin-wave contributions to nuclear magnetic relaxation in magnetic metals}
\author{V.Yu. Irkhin$^{*}$ and M.I. Katsnelson}
\address{Institute of Metal Physics, 620219 Ekaterinburg, Russia}
\maketitle

\begin{abstract}
The longitudinal and transverse nuclear magnetic relaxation rates $1/T_1(T)$
and $1/T_2(T)$ are calculated for three- and two-dimensional ($3D$ and $2D$)
metallic ferro- and antiferromagnets (FM and AFM) with localized magnetic
moments in the spin-wave temperature region. The contribution of the
one-magnon decay processes is strongly enhanced in comparison with the
standard $T$-linear Korringa term, especially for the FM case. For the $3D$
AFM case this contribution diverges logarithmically, the divergence being
cut at the magnon gap $\omega _0$ due to magnetic anisotropy, and for the
$2D$ AFM case as $\omega _0^{-1}$. The electron-magnon scattering processes
yield $T^2\ln (T/\omega _0)$ and $T^2/\omega _0^{1/2}$-terms in $1/T_1$ for
the $3D$ AFM and $2D$ FM cases, respectively. The two-magnon (``Raman'')
contributions are investigated and demonstrated to be large in the $2D$ FM
case. Peculiarities of the isotropic $2D$ limit (where the correlation
length is very large) are analyzed.
\end{abstract}


\section{Introduction}

Nuclear magnetic resonance (NMR), which is one of most powerful tools for
investigating various physical properties, has a number of peculiarities for
magnetically ordered materials. Last time, a number of new classes of
magnets have been studied by this method, e.g., heavy-fermion compounds \cite
{hf}, ferromagnetic films and monolayers \cite{films}, low-dimensional
systems including copper-oxide perovskites \cite{Mehr}, etc. Thus the
problem of theoretical description of various NMR characteristics of magnets
is topical again. This problem was already a subject of great interest since
50-60s when the interaction of nuclear magnetic moments with spin waves in
localized-spin Heisenberg model was studied \cite{Mor56,Turov}. However,
this model is inadequate to describe the most interesting systems mentioned
above where the role of conduction electrons is essential in magnetic
properties. Usually the data on the longitudinal nuclear magnetic relaxation
rate $1/T_1$ (this NMR characteristic is probably most convenient to compare
experimental results with theoretical predictions) are discussed within
itinerant-electron models such as Hubbard model or phenomenological
spin-fluctuation theories. Ueda and Moriya \cite{UM,Mor} calculated the
dependences $1/T_1(T)$ for weak itinerant magnets, especial attention being
paid to the paramagnetic region, and obtained strong temperature effects.
Later this approach was extended to the two-dimensional case and extensively
developed in connection with high-$T_c$ superconductors and related
compounds \cite{Mor1,Pines}. On the other hand, in a number of systems
(e.g., in most rare-earth compounds which are also a subject of NMR
investigations, see, e.g., Refs.\cite{re}) the $s-d(f)$ exchange model with
well-separated localized and itinerant magnetic subsystems is more adequate.
Magnetic properties in such a situation differ essentially from those in the
paramagnon regime (see, e.g., discussion in Refs.\cite{IKJP90,afm}). At the
same time, the contributions to nuclear magnetic relaxation rate owing to
electron-magnon interaction are not investigated in detail.

In the present work we obtain the dependences of $1/T_1(T)$ and the
linewidth $1/T_2(T)$ in the spin-wave region for three- and two-dimensional (%
$3D$ and $2D$) metallic magnets with well-defined local magnetic moments. In
Sect.2 we discuss the general formalism and physical picture of hyperfine
interactions. In Sects.3 and 4 we calculate various contributions to the
relaxation rates in metallic ferro- and antiferromagnets. In Sect.5 we
analyze the isotropic $2D$ case where at finite temperatures the long-range
order is absent, but the correlation length is very large.

\section{Hyperfine interactions}

We start from the standard Hamiltonian of the hyperfine interaction \cite
{abragam}
\begin{equation}
H_{hf}={\bf hI,\,}h_\alpha =A_{\alpha \beta }S_\beta
\end{equation}
$\widehat{A}$ being the hyperfine interaction matrix, which contains the
Fermi (contact) and dipole-dipole contributions,
\begin{equation}
A_{\alpha \beta }=A^F\delta _{\alpha \beta }+A_{\alpha \beta }^{dip}.
\end{equation}
The Fermi hyperfine interaction is proportional to the electron density at
the nucleus and therefore only $s$-states participate in it, the
contribution of core $s$-states (which are polarized due to local magnetic
moments) being much larger than of conduction electrons. It is just the
consequence of considerably smaller localization area (and therefore higher
density on nuclei) for the core states.

The dipole contribution to $H_{hf}$ can be represented as \cite{abragam}
\begin{equation}
H_{hf}^{dip}=\frac a2\{[\frac 16(I^{+}S^{-}+I^{-}S^{+})-\frac 13%
I^zS^z]F^{(0)}+I^{+}S^{+}F^{(2)}+2(I^zS^{+}+I^{+}S^z)F^{(1)}\}+h.c\text{.}
\label{hdip}
\end{equation}
where
\begin{eqnarray}
F^{(0)} &=&\langle (1-3\cos ^2\theta )/r^3\rangle ,F^{(1)}=\langle \sin
\theta \cos \theta \exp (-i\phi )/r^3\rangle ,  \nonumber \\
F^{(2)} &=&\langle \sin ^2\theta \exp (-2i\phi )/r^3\rangle ,a=-\frac 32%
\gamma _e\gamma _n
\end{eqnarray}
Here $\langle ...\rangle $ is the average over the electron subsystem
states, $\gamma _e$ and $\gamma _n$ are gyromagnetic ratios for electron and
nuclear moments, respectively. In the case of the {\it local} cubic symmetry
we have $F^{(a)}=0.$ It is obvious that magnetic $f$- or $d$-electrons
dominate also in dipole interactions because of large spin polarization.
Hence the direct interaction of nuclear spins with that of conduction
electrons can be neglected in magnets with well-defined local magnetic
moments. Nevertheless, conduction electrons do effect nuclear relaxation via
their influence on the local-moment system; besides that, as we shall see
below, such contributions possess large exchange enhancement factors. The
investigation of these effects is one of the main aims of this work.

A general way to consider all these contributions is using the Green's
function method which leads to the following expression for the longitudinal
nuclear magnetic relaxation rate \cite{Mor63}
\begin{eqnarray}
\frac 1{T_1} &=&-\frac T{2\pi }\text{Im}\sum_{{\bf q}}\langle \langle h_{%
{\bf q}}^{+}|h_{-{\bf q}}^{-}\rangle \rangle _{\omega _n}/\omega _n,
\label{tt1} \\
\frac 1{T_2} &=&\frac 1{2T_1}-\frac T{2\pi }\lim_{\omega \rightarrow 0}\text{%
Im}\sum_{{\bf q}}\langle \langle h_{{\bf q}}^z|h_{-{\bf q}}^z\rangle \rangle
_\omega /\omega  \label{tt2}
\end{eqnarray}
($\omega _n=\langle h^z\rangle \ll T$ is the NMR frequency). As follows from
(\ref{hdip}),
\begin{eqnarray}
h^{-} &=&(A^F+\frac 13aF^{(0)})S^{-}+aF^{(2)}S^{+}+2aF^{(1)}S^z, \\
h^z &=&(A^F-\frac 23aF^{(0)})S^z+a(F^{(1)}S^{+}+aF^{(1)*}S^{-})
\end{eqnarray}
Then we derive
\begin{eqnarray}
\frac 1{T_1} &=&\frac T2\{[(A^F+\frac 13aF^{(0)})^2+a^2|F^{(2)}|^2]K^{+-}
\nonumber \\
&&\ \ \ \ \ \ \ +2a(A^F+\frac 13aF^{(0)})\text{Re}%
F^{(2)}K^{++}+4a^2|F^{(1)}|^2K^{zz}\}  \label{FK}
\end{eqnarray}
\begin{equation}
\frac 1{T_2}=\frac 1{2T_1}+\frac T2\{(A^F-\frac 23%
aF^{(0)})^2K^{zz}+a^2[2|F^{(1)}|^2K^{+-}+(F^{(1)})^2K^{++}+(F^{(1)*})^2K^{--}]\}
\label{FK1}
\end{equation}
where the quantities $K^{\alpha \beta }$ are defined by
\begin{equation}
K^{\alpha \beta }=-\frac 1\pi \lim_{\omega \rightarrow 0}\text{Im}\sum_{{\bf %
q}}\langle \langle S_{{\bf q}}^\alpha |S_{-{\bf q}}^\beta \rangle \rangle
_\omega /\omega
\end{equation}
As we shall see below in Sect.5, $\omega _n\neq 0$ which enters (\ref{FK})
(but not the second term of (\ref{FK1})) may become important in the case of
very small magnetic anisotropy.

Formula (\ref{tt1}) has a rather general character. On the other hand, the
problem of calculating the NMR linewidth is much more complicated. The
Moriya formula (\ref{tt2}) is in fact applicable only in the case where the
line has the Lorentz form (i.e., charactersitic frequency of hyperfine field
fluctuations is large in comparison with their amplitude) \cite{abragam}. In
insulating crystals the latter condition is usually violated, the lineform
being close to Gaussian with the width determined by dipole interactions of
nuclear spins. At the same time, in metals the Korringa relaxation described
by the formula (\ref{tt2}) usually dominates, so that we will use this. A
peculiar case is provided by conducting systems which are on the borderline
of ferro- or antiferromagnetic instability (i.e. with large correlation
length $\xi $), e.g., copper-oxide superconductors \cite{thelen}. Under this
condition, the anisotropic Ruderman-Kittel interaction between nuclear spins
turns out to be greatly ehanced and dominates over the dipole interaction.
The lineform turns out to be Gaussian with the width being estimated as

\begin{equation}
\left( \frac 1{T_2}\right) ^2\propto A^2\sum_{{\bf q}}\chi ^2({\bf q},\omega
=0)  \label{gauss}
\end{equation}
where $\chi ({\bf q},\omega )$ is the dynamical spin susceptibility of
electron system. We will use this result in Sect.5 when discussing 2D
systems which do possess large correlation length.

\section{Ferromagnetic metals}

We proceed with the $s-d(f)$ exchange model Hamiltonian
\begin{equation}
\ H=\sum_{{\bf k}\sigma }t_{{\bf k}}c_{{\bf k}\sigma }^{\dagger }c_{{\bf k}%
\sigma }-I\sum_{i\alpha \beta }{\bf S}_i\mbox {\boldmath $\sigma $}_{\alpha
\beta }c_{i\alpha }^{\dagger }c_{i\beta }+\sum_{{\bf q}}J_{{\bf q}}{\bf S}_{%
{\bf -q}}{\bf S}_{{\bf q}}+H_a  \label{H}
\end{equation}
where $t_{{\bf k}}$ is the band energy, ${\bf S}_i$ and ${\bf S}_{{\bf q}}$
are spin-density operators and their Fourier transforms, $\sigma $ are the
Pauli matrices, $H_a$ is the anisotropy Hamiltonian which results in
occurrence of the gap $\omega _0$ in the spin-wave spectrum. For convenience
we include explicitly in the Hamiltonian the Heisenberg exchange interaction
with the parameters $J_{{\bf q}},$ although really this may be, e.g., the
Ruderman-Kittel-Kasuya-Yosida (RKKY) interaction. It should be noted that
similar results may be reproduced for the localized-moment Hubbard magnets
(cf.\cite{IKJP90,Ent}).

First we consider the ferromagnetic (FM) case. Then $K^{++}=0$ and the
relaxation rates (\ref{FK}), (\ref{FK1}) are the sums of transverse $%
(\propto K^{+-})$ and longitudinal $(\propto K^{zz})$ terms. Passing to the
magnon representation we obtain
\begin{equation}
\langle \langle S_{{\bf q}}^{+}|S_{-{\bf q}}^{-}\rangle \rangle _\omega
=2S/[\omega -\omega _{{\bf q}}+i\gamma _{{\bf q}}(\omega )]  \label{fgf}
\end{equation}
where $\omega _{{\bf q}}=2S(J_{{\bf q}}-J_0)+\omega _0$ is the magnon
frequency, $\gamma _{{\bf q}}(\omega )\propto \omega $ is the magnon
damping. Then we have
\begin{equation}
K^{+-}=2S\sum_{{\bf q}}\frac{\gamma _{{\bf q}}(\omega _n)}{\pi \omega
_n\omega _{{\bf q}}^2}  \label{gfm}
\end{equation}
(cf. Refs.\cite{IKJP90,UFN}). The damping in the denominator of (\ref{gfm})
can be neglected for both localized-moment and itinerant-electron magnets
(in the latter case the expression (\ref{fgf}) corresponds to the RPA
structure, see Ref.\cite{IKJP90}) due to smallness of $\omega _n.$ On the
contrary, temperature dependences of magnetization, resistivity etc. in weak
itinerant magnets are just determined by the damping in the denominator,
i.e. by paramagnon excitations rather than by spin waves \cite{Mor}.

The damping owing to the one-magnon decay processes is given by the
well-known expression
\begin{eqnarray}
\gamma _{{\bf q}}^{(1)}(\omega ) &=&-2\pi I^2S\sum_{{\bf k}}(n_{{\bf %
k\uparrow }}-n_{{\bf k-q\downarrow }})  \label{G1F} \\
\ \times \delta (\omega +t_{{\bf k\uparrow }}-t_{{\bf k-q\downarrow }})
&\simeq &2\pi I^2S\omega \lambda _{{\bf q}}  \nonumber
\end{eqnarray}
where $t_{{\bf k}\sigma }=$ $t_{{\bf k}}-\sigma IS,\,t_{{\bf k}}$ is
referred to the Fermi level, $n_{{\bf k}\sigma }=n(t_{{\bf k}\sigma })$ is
the Fermi function,
\begin{equation}
\lambda _{{\bf q}}=\sum_{{\bf k}}\delta (t_{{\bf k\uparrow }})\delta (t_{%
{\bf k-q\downarrow }}).
\end{equation}
The linearity of spin fluctuation damping in $\omega $ is the characteristic
property of metals. According to (\ref{FK}) this leads to $T$-linear
contributions to $1/T_1$ which is the Korringa law \cite{Kor}. Note that the
simplest expression for the Korringa relaxation

\begin{equation}
1/T_1\simeq 1/T_2\simeq A^2\rho _{\uparrow }\rho _{\downarrow }T,
\label{kor}
\end{equation}
where $A$ is an effective hyperfine interaction constant, $\rho _\sigma $
are the partial densities of electron states at the Fermi level, is
practically never applicable for magnetic metals (e.g., exchange enhancement
factors can change even the order of magnitude of $1/T_1$ \cite{Mor,UFN}).
Accurate expression for the ``Korringa'' contribution in the case under
consideration can be derived by the substitution (\ref{gfm}) and (\ref{G1F})
into (\ref{FK}).

The damping (\ref{G1F}) has the threshold value of $q,$ which is determined
by the spin splitting $\Delta =2|I|S$, $q^{*}=\Delta /v_F$ ($v_F$ is the
electron velocity at the Fermi level). The quantity $q^{*}$ determines a
characteristic temperature and energy scale
\begin{equation}
\omega ^{*}=\omega (q^{*})={\cal D}(q^{*})^2\sim (\Delta /v_F)^2T_C
\label{T*}
\end{equation}
with ${\cal D}$ the spin-wave stiffness.

Besides $3D$ magnets, consideration of the $2D$ case is of interest (this
may be relevant, e.g., for layered magnets and ferromagnetic films; for more
details see Sect.5). We have
\begin{equation}
\lambda _{{\bf q}}=\theta (q-q^{*})\times \left\{
\begin{tabular}{ll}
$(qv_F)^{-1},$ & $D=3$ \\
$\frac 1\pi (q^2v_F^2-\Delta ^2)^{-1/2},$ & $D=2$%
\end{tabular}
\right.
\end{equation}
After integration for the parabolic electron spectrum ($q^{*}$ plays the
role of the lower cutoff), the one-magnon damping contribution to (\ref{gfm}%
) takes the form
\begin{equation}
\delta ^{(1)}K^{+-}=\frac{\rho _{\uparrow }\rho _{\downarrow }}{{\cal D}^2m^2%
}\times \left\{
\begin{tabular}{ll}
$1/4,$ & $D=3$ \\
$1/(\pi q^{*}),$ & $D=2$%
\end{tabular}
\right.  \label{t1f}
\end{equation}
with
\begin{equation}
\rho _\sigma =\frac{m\Omega _0}{2\pi }\times \left\{
\begin{tabular}{ll}
$k_{F\sigma }/\pi ,$ & $D=3$ \\
$1,$ & $D=2$%
\end{tabular}
\right.
\end{equation}
$m$ the electron effective mass, $\Omega _0$ the lattice cell volume (area).
Thus in the $3D$ case the factor of $I^2$ is canceled, and the factor of $%
I^{-1}$ occurs in the $2D$ case and we obtain a strongly enhanced $T$-linear
Korringa-type term (remember that ${\cal D}\sim J\sim I^2\rho $ for the RKKY
interaction). This means that the contribution of conduction electrons to $T$%
-linear relaxation rate via their interaction with localized spins is indeed
much more important than the ``direct'' contribution (\ref{kor}):
perturbation theory in the $s-d$ exchange coupling parameter $I$ turns out
to be singular. Earlier such contributions (for the $3D$ case) were
calculated by Weger \cite{Weg} and Moriya \cite{Mor64} for iron-group
metals. However, Moriya has concluded that for these materials they are not
important in comparison with orbital current contributions. In the case
under consideration (where magnetic subsystem is well separated from the
conductivity electrons) the situation is different and the spin-wave
contribution in $1/T_1$ is normally the most important.

The one-magnon decay contribution (\ref{t1f}) is absent for so-called
half-metallic ferromagnets, e.g., some Heusler alloys, where electron states
with one spin projection only are presented at the Fermi surface \cite
{HFM,UFN}. In such a situation we have to consider two-magnon scattering
processes. In this connection, it is worthwhile to note an important
difference between relaxation processes via phonons and via magnons. The
main difference is due to the gap in magnon spectrum. Usually $\omega
_0>\omega _n$ and therefore one-magnon processes contribute to the
relaxation rate due to magnon damping only (cf. discussion of the
phonon-induced relaxation processes in Ref.\cite{abragam}). However, the
mechanisms of magnon damping in magnetic dielectrics (magnon-magnon
interactions) are different from those in magnetic metals and degenerate
semiconductors \cite{Aus,afm}.

The damping in a conducting ferromagnet owing to electron-magnon
(two-magnon) scattering processes is calculated in Refs.\cite{Aus,IKJP90}
and has the form
\begin{eqnarray}
\frac{\gamma _{{\bf q}}^{(2)}(\omega )}\omega &=&\pi I^2\sum_{{\bf kp}\sigma
}\left( \frac{t_{{\bf k+q}}-t_{{\bf k}}}{t_{{\bf k+q}}-t_{{\bf k}}+2\sigma IS%
}\right) ^2(\omega _{{\bf p}}-\omega )\frac{\partial n_{{\bf k}\sigma }}{%
\partial t_{{\bf k}}}\frac{\partial N_{{\bf p}}}{\partial \omega _{{\bf p}}}
\nonumber  \\
&&\ \ \ \ \ \ \ \ \ \ \ \ \ \ \ \times \delta (t_{{\bf k}}-t_{{\bf k-p+q}})
\label{Gam2gen}
\end{eqnarray}
where $N_{{\bf p}}=N(\omega _{{\bf p}})$ is the Bose function. Substituting
this into (\ref{gfm}) and performing integration we obtain for $D=3$%
\begin{equation}
\delta ^{(2)}K^{+-}=\frac{\Omega _0T^{1/2}}{128\pi ^2Sm^2{\cal D}^{7/2}}%
\sum_\sigma \rho _\sigma ^2\times \left\{
\begin{tabular}{ll}
$3\pi ^{1/2}\zeta (\frac 32)T,$ & $T\ll \omega ^{*}$ \\
$8M_3\omega ^{*},$ & $T\gg \omega ^{*}$%
\end{tabular}
\right.  \label{f2}
\end{equation}
where $\zeta (z)$ is the Riemann function,
\begin{equation}
M_3=\int_0^\infty dx\left[ \frac 1{x^2}-\frac{x^2\exp x^2}{(\exp x^2-1)^2}%
\right] \simeq 0.65
\end{equation}
The contribution (\ref{f2}) should play the dominant role in the
half-metallic ferromagnets \cite{UFN}. Besides that, this contribution may
modify considerably the temperature dependence of $1/T_1$ in ``usual''
ferromagnets, a crossover from $T^{5/2}$ to $T^{3/2}$ dependence of the
correction taking place.

For $D=2$ at $T,\omega ^{*}\gg \omega _0$ small magnon momenta of order of $%
\left( \omega _0/{\cal D}\right) ^{1/2}$ make the main contribution to (\ref
{gfm}). To calculate the integral one can use the high-temperature
expression for $N_{{\bf p}}=T/\omega _{{\bf p}}$. As a result, one gets
\begin{equation}
\delta ^{(2)}K^{+-}=\frac{\Omega _0^3k_FM_2}{8\pi ^4S{\cal D}^{5/2}\omega
_0^{1/2}}T  \label{f22}
\end{equation}
with
\begin{eqnarray}
M_2 &=&\int_0^\infty \frac{dx}{1+x^2}\int\limits_0^{\pi /2}\frac{d\varphi
\sin ^2\varphi }{\left( \sin ^2\varphi +x^2\right) ^{3/2}}  \nonumber \\
\ &=&\int_0^\infty dy\left[ 1+\frac{y^2}{\sqrt{1+y^2}}\ln \frac{\sqrt{1+y^2}%
-1}y\right] \simeq 1.23
\end{eqnarray}
Thus in the $2D$ FM case, in contrast with $3D$ one, the relaxation rate $%
1/T_1$ is strongly dependent on the anisotropy gap.

Consider now the second term in the tranverse relaxation rate $1/T_2(T)$ (%
\ref{FK1}), which is normally determined by $K^{zz},$ and the longitudinal
contribution to relaxation rate $1/T_1$ in (\ref{FK}), which is due to
dipole-dipole interactions with the characteristic constant $\widetilde{A}%
\sim a|F^{(1)}|$. The simplest calculation from the longitudinal Green's
function for the localized-spin subsystem gives
\begin{eqnarray}
\langle \langle S_{{\bf q}}^z|S_{-{\bf q}}^z\rangle \rangle _\omega &=&\sum_{%
{\bf p}}\frac{N_{{\bf p}}-N_{{\bf p-q}}}{\omega -\omega _{{\bf p-q}}+\omega
_{{\bf p}}}, \\
K^{zz} &=&\sum_{{\bf qp}}\left( -\frac{\partial N_{{\bf p}}}{\partial \omega
_{{\bf p}}}\right) \delta (\omega _{{\bf q}}-\omega _{{\bf p}}).  \label{kzz}
\end{eqnarray}
The quantity (\ref{kzz}) has been considered in Refs.\cite{Rob,Turov} as a
contribution to the NMR line width $1/T_2.$ The integration in the $3D$ case
gives the logarithmic singularity
\begin{equation}
K^{zz}=\frac{\Omega _0^2}{16\pi ^4{\cal D}^3}T\ln \frac T{\omega _0}
\label{zz3d}
\end{equation}
For $D=2$ this singular term is inversely proportional to the magnetic
anisotropy parameter and very large:
\begin{equation}
K^{zz}=\left( \frac{\Omega _0}{4\pi {\cal D}}\right) ^2N(\omega _0)\simeq
\left( \frac{\Omega _0}{4\pi {\cal D}}\right) ^2\frac T{\omega _0},T\gg
\omega _0  \label{zz2d}
\end{equation}
For small enough $\omega _0$ and $\widetilde{A}\sim A$ this contribution can
dominate over the ``Korringa'' contribution (\ref{t1f}) in $1/T_1$ at $%
T>\omega _0/|I\rho |$. The leading contribution to $K^{zz}$ from the $s-d$
interaction is determined by
\begin{eqnarray}
\delta \langle \langle S_{{\bf q}}^z|S_{-{\bf q}}^z\rangle \rangle _\omega
&=&2I^2S\sum_{{\bf kp}\sigma }\frac 1{(\sigma \omega +\omega _{{\bf p-q}%
}-\omega _{{\bf q}})^2}  \nonumber \\
&&\ \ \ \ \ \ \ \ \ \ \ \ \ \ \ \ \ \ \ \ \ \ \ \ \ \times \frac{n_{{\bf %
k\downarrow }}(1-n_{{\bf k+p-q\uparrow }})+N_{{\bf p}}(n_{{\bf k\downarrow }%
}-n_{{\bf k+p-q\uparrow }})}{t_{{\bf k\downarrow }}-t_{{\bf k+p-q\uparrow }%
}+\sigma \omega -\omega _{{\bf p}}}
\end{eqnarray}
However, it is not singular in $\omega _0$ and practically never important.

\section{Antiferromagnetic metals}

Now we consider the spiral antiferromagnetic (AFM) structure along the $x$%
-axis with the wavevector {\bf Q \ }
\[
\langle S_i^z\rangle =S\cos {\bf QR}_i,\,\langle \,S_i^y\rangle =S\sin {\bf %
QR}_i,\,\langle S_i^x\rangle =0
\]
We introduce the local coordinate system
\begin{eqnarray*}
S_i^z &=&\hat S_i^z\cos {\bf QR}_i-\hat S_i^y\sin {\bf QR}_i, \\
\,S_i^y &=&\hat S_i^y\cos {\bf QR}_i+\hat S_i^z\sin {\bf QR}_i,\,S_i^x=\hat S%
_i^x
\end{eqnarray*}
Further we pass from spin operators ${\bf \hat S}_i$ to the spin deviation
operators $b_i^{\dagger },b_i$ and, by the canonical transformation $b_{{\bf %
q}}^{\dagger }=u_{{\bf q}}\beta _{{\bf q}}^{\dagger }-v_{{\bf q}}\beta _{-%
{\bf q}},$ to the magnon operators. Hereafter we consider for simplicity
two-sublattice AFM ordering (2{\bf Q }is equal to a reciprocal lattice
vector, so that $\cos ^2{\bf QR}_i=1,\,\sin ^2{\bf QR}_i=0$).

Calculating the Green's functions to second order in $I$ (to second order in
the formal quasiclassical parameter $1/2S,$ cf. Refs.\cite{afm,kondo}) we
derive
\begin{eqnarray}
\langle \langle b_{{\bf q}}|b_{{\bf q}}^{\dagger }\rangle \rangle _\omega &=&%
\frac{\omega +C_{{\bf q-}\omega }}{(\omega -C_{{\bf q}\omega })(\omega +C_{%
{\bf q-}\omega })+D_{{\bf q}\omega }^2}  \label{FG} \\
\langle \langle b_{-{\bf q}}^{\dagger }|b_{{\bf q}}^{\dagger }\rangle
\rangle _\omega &=&\frac{D_{{\bf q}\omega }}{(\omega -C_{{\bf q}\omega
})(\omega +C_{{\bf q-}\omega })+D_{{\bf q}\omega }^2}  \label{FGa}
\end{eqnarray}
with
\begin{eqnarray}
C_{{\bf q}\omega } &=&S(J_{{\bf Q+q},\omega }^{tot}+J_{{\bf q}\omega
}^{tot}-2J_{{\bf Q}0}^{tot})+\sum_{{\bf p}}[C_{{\bf p}}\Phi _{{\bf pq}\omega
}  \nonumber \\
&&\ \ \ \ \ \ \ \ \ \ \ \ \ \ \ \ \ \ \ \ -(C_{{\bf p}}-D_{{\bf p}})\Phi _{%
{\bf p}00}+\phi _{{\bf pq}\omega }^{+}+\phi _{{\bf pq}\omega }^{-}]+g_{{\bf q%
}}  \label{CD} \\
D_{{\bf q}\omega } &=&D_{{\bf q-}\omega }=S(J_{{\bf q}\omega }^{tot}-J_{{\bf %
Q+q},\omega }^{tot})+\sum_{{\bf p}}D_{{\bf p}}\Phi _{{\bf pq}\omega }+h_{%
{\bf q}}  \nonumber
\end{eqnarray}
The $s-d$ exchange contributions of the first order in $1/2S$ correspond to
the RKKY approximation
\begin{equation}
J_{{\bf q}\omega }^{tot}=J_{{\bf q}}+I^2\sum_{{\bf k}}\frac{n_{{\bf k}}-n_{%
{\bf k-q}}}{\omega +t_{{\bf k}}-t_{{\bf k-q}}}  \label{RKKY}
\end{equation}
($n_{{\bf k}}=n(t_{{\bf k}})$ is the Fermi function), the second summand in (%
\ref{RKKY}) being the $\omega $-dependent RKKY indirect exchange
interaction. The function $\Phi $, which determines the second-order
corrections, is given by
\begin{eqnarray}
\Phi _{{\bf pq}\omega } &=&(\phi _{{\bf pq}\omega }^{+}-\phi _{{\bf pq}%
\omega }^{-})/\omega _{{\bf p}},  \label{Fi} \\
\phi _{{\bf pq}\omega }^{\pm } &=&I^2\sum_{{\bf k}}\frac{n_{{\bf k}}(1-n_{%
{\bf k+p-q}})+N(\pm \omega _{{\bf p}})(n_{{\bf k}}-n_{{\bf k+p-q}})}{\omega
+t_{{\bf k}}-t_{{\bf k+p-q}}\mp \omega _{{\bf p}}}  \nonumber
\end{eqnarray}
where
\[
\omega _{{\bf p}}=(C_{{\bf p}}^2-D_{{\bf p}}^2)^{1/2}=[4S^2(J_{{\bf p}}-J_{%
{\bf Q}})(J_{{\bf Q+p}}-J_{{\bf Q}})+\omega _0^2]^{1/2}
\]
is the magnon frequency to zeroth order in $I\,$and $1/2S$. The $\omega $%
-independent corrections $g_{{\bf q}},h_{{\bf q}}$ that describe the
``direct'' magnon-magnon interaction are written down in Refs.\cite
{afm,kondo}.

Now we consider the effects of electron-magnon interaction. The intrasubband
one-magnon damping (which is absent in the FM case) is finite at arbitrarily
small $q$ \cite{IKFMM}. Similar to the FM case, the contributions of
intersubband transitions (which correspond to small $|{\bf q-Q}|$) are cut
at the characteristic temperature and energy scale
\begin{equation}
\omega ^{*}=\omega (q^{*})=cq^{*}\sim (\Delta /v_F)T_N,  \label{T**}
\end{equation}
We have
\begin{equation}
K^{+-}=-\frac{2S}\pi \lim_{\omega \rightarrow 0}\text{Im}\sum_{{\bf q}%
}\omega ^{-1}C_{{\bf q}\omega }/\omega _{{\bf q}}^2,
\end{equation}
and the term with $K^{++}$ in (\ref{FK}) vanishes due to the property $D_{%
{\bf q}\omega }=-D_{{\bf q+Q}\omega }.$ The one-magnon contribution owing to
the imaginary part of (\ref{RKKY}) in the 3D case takes after integration
the form
\begin{equation}
\delta ^{(1)}K^{+-}=\frac{S^2\Omega _0}{\pi ^2c^2}\left( P_0\ln \frac{\omega
_{\max }}{\omega _0}+P_{{\bf Q}}\ln \frac{\omega _{\max }}{\omega _0^{*}}%
\right) .  \label{K1}
\end{equation}
Here $c$ is the magnon velocity defined by $\omega _{{\bf p}}^2=\omega _{%
{\bf p+Q}}^2=\omega _0^2+c^2p^2$,
\begin{equation}
P_{{\bf p}}=I^2\lim_{{\bf q}\rightarrow 0}|{\bf q-p}|\sum_{{\bf k}}\delta
(t_{{\bf k}})\delta (t_{{\bf k-q+p}}),
\end{equation}
(the quantity $P_0$ depends, generally speaking, on the direction of the
vector ${\bf q,}$ see Refs.\cite{Mor,Pi,Kal}), the second logarithm in the
brackets of (\ref{K1}) contains the cutoff
\[
\omega _0^{*}=\sqrt{\omega _0^2+(\omega ^{*})^2}
\]
The ``enhancement'' factor in (\ref{K1}) is smaller than in the FM case
because of the linear dispersion law of magnons, but this contribution still
dominates over the ``usual'' Korringa term (\ref{kor}). Besides that, a
large logarithmic factor occurs (in the isotropic case, this is cut at $%
\omega _n$ only). Note that a similar logarithmic singularity in $1/T_1$
takes place for $3D$ itinerant-electron antiferromagnets \cite{UM}. It is
interesting that the intersubband contribution does not lead here to
enhancing the singularity, unlike the situation for the magnon damping,
magnetic and transport properties\cite{IK95,afm}. Under the ``nesting''
conditions ($t_{{\bf k+Q}}\simeq -t_{{\bf k}}$ in a large part of the Fermi
surface) the singularity is not enhanced as well.

The singularity becomes stronger in the $2D$ case where integration gives
\begin{equation}
\delta ^{(1)}K^{+-}=\frac{S^2\Omega _0}{\pi c\omega _0}\left( \frac \pi 2%
P_0+P_{{\bf Q}}\arctan \frac{\omega _0}{\omega ^{*}}\right)  \label{a12}
\end{equation}
This fact may be important when treating experimental data on layered AFM
metals.

The contribution owing to electron-magnon scattering processes is determined
by the imaginary part of the function (\ref{Fi}). After a little
manipulation we obtain
\begin{equation}
\delta ^{(2)}K^{+-}\simeq 2SL\sum_{{\bf p}\rightarrow 0,{\bf q}}\frac 1{%
q\omega _{{\bf q+p}}^2}\left( -\frac{\partial N_{{\bf p}}}{\partial \omega _{%
{\bf p}}}\right) [P_0+P_{{\bf Q}}\widetilde{\phi }(q)]
\end{equation}
where $L=2S(J_0-J_{{\bf Q}}),\widetilde{\phi }(q<q^{*})=0,\,\widetilde{\phi }%
(q\gg q^{*})=1.$ The integration in the $3D$ case yields
\begin{equation}
\delta ^{(2)}K^{+-}=\frac{SL\Omega _0^2}{8\pi ^4c^4}\left[ P_0f(T,\omega
_0)+P_{{\bf Q}}f(T,\omega _0^{*})\right]
\end{equation}
where
\begin{eqnarray}
f(T,\omega _0) &=&\int_{\omega _0}^\infty d\omega \omega \left( -\frac{%
\partial N(\omega )}{\partial \omega }\right) \ln \frac{\omega _{\max }}%
\omega  \nonumber \\
\ &\simeq &T\ln \frac T{\omega _0}\left( \ln \frac{\omega _{\max }}{\omega _0%
}-\frac 12\ln \frac T{\omega _0}\right) ,T\gg \omega _0
\end{eqnarray}
Thus we have $1/T_1\propto T^2\ln T.$ In the $2D$ case we derive
\begin{equation}
\delta ^{(2)}K^{+-}\simeq T\frac{SL\Omega _0^2}{4\pi ^4c^4}\left( P_0\ln ^2%
\frac T{\omega _0}+P_{{\bf Q}}\ln ^2\frac T{\omega _0^{*}}\right) ,
\label{a22}
\end{equation}
so that the singularity is not enhanced in comparison with the $3D$ case.

The contributions owing to longitudinal fluctuations will be estimated for
the localized subsystem only. We obtain
\begin{equation}
K^{zz}\simeq \sum_{{\bf pq}}\frac{L^2}{2\omega _{{\bf p}}^2}\left( -\frac{%
\partial N_{{\bf p}}}{\partial \omega _{{\bf p}}}\right) \delta (\omega _{%
{\bf q}}-\omega _{{\bf p}})  \label{kzza}
\end{equation}
The corresponding contribution to $1/T_2$ was considered in Ref.\cite{Mor56}%
. The term in the longitudinal relaxation rate determined by (\ref{kzza}) is
estimated as
\begin{equation}
\delta ^{(z)}(1/T_1)\propto \widetilde{A}^2\times \left\{
\begin{tabular}{ll}
$T^3/J^4,$ & $D=3$ \\
$T^2/J^3,$ & $D=2$%
\end{tabular}
\right.
\end{equation}
Provided that the dipole-dipole contributions in (\ref{FK}) are considerable
($\widetilde{A}\sim A$), this term can dominate over the ``Korringa'' term (%
\ref{K1}) of order of $A^2I^2\rho ^2T\ln |J/\omega _0|/J^2$ at $T/|J|>|I\rho
|\ln ^{1/2}|J/\omega _0|$ only. Note that this two-magnon contribution is
similar to the two-phonon (Raman) contribution in the spin-lattice
relaxation. The existence of the gap $\omega _0$ is not important here (at
least if it is sufficiently small), but the matrix elements of interaction
of nuclear spins with magnons are singular, unlike those for acoustic
phonons ($|M_{{\bf q}\rightarrow 0}|^2\sim 1/q$ instead of $q$ ). Therefore
we have a $T^3$ law instead of $T^7$ one for the phonon scattering \cite
{abragam}.

\section{Isotropic 2D case and NMR in layered and frustrated magnets}

Now we investigate the case of layered magnets, in particular, the isotropic
$2D$ limit. A detailed treatment of the spin correlation functions and
corresponding spin-fluctuation contributions to $1/T_1$ in the isotropic $2D$
Heisenberg antiferromagnets with $\omega _n\rightarrow 0$ was performed in
Ref.\cite{t13}. Here we calculate also corrections owing to electron-magnon
interaction.

The magnetic ordering temperature is determined by magnetic anisotropy or
interlayer coupling,
\begin{equation}
T_M\sim |J|S^2/\ln (|J|S^2/\max \{\omega _0,|J^{\prime }|\})  \label{tm}
\end{equation}
(for more details see, e.g., Refs.\cite{SSWT}). Despite the absence of the
long-range ordering (LRO) at finite temperatures, spin-wave description
holds even in the pure $2D$ isotropic case in the broad temperature region
up to $T\sim |J|S$ (i.e., $T_M\rightarrow |J|S^2$) owing to strong
short-range order (SRO). In a more general case of finite $\omega _0$ and $%
J^{\prime }$, this description holds at $T\gg T_M.$ A possibilty to describe
LRO without introducing anomalous averages (like sublattice magnetization)
in terms of singularities of the spin correlation function was demonstrated
in Ref.\cite{IKZ}. Such an approach enables one to obtain an unified
description of ordered and disordered phases. In the pure $2D$ case the
magnetization (or sublattice magnetization for the AFM case) $\overline{S}$
is replaced in both magnetic and electronic properties by the square root of
the Ornstein-Cernike peak (see, e.g., \cite{IKJP91}). The gap in the
effective spin-wave spectrum appears at finite temperatures, which is
determined by the inverse correlation length. The correlation length in the
situation under consideration is estimated as \cite{Arovas}
\begin{equation}
\xi \propto \exp \left( \pi |J|S^2/T\right) \text{\ }  \label{korr}
\end{equation}
As shows the two-loop scaling theory\cite{Chakraverty}, the preexponentional
factor is temperature independent; quantum effects can renormalize the
exchange parameter $J$ \cite{SSWT}.

To describe formally NMR in the absence of LRO ($\langle {\bf S}_i\rangle =0$%
) we follow to Ref.\cite{IKZ} and consider the autocorrelation function of
the nuclear spin ${\bf I}$ \cite{Mori}. Performing calculations with the
simplest Hamiltonian $H_{hf}=A{\bf IS}_i$ to second order in $A$ we derive
\[
(I^{+},I^{-})^\omega =\frac 23I(I+1)[-i\omega +\Sigma (\omega )]
\]
with the memory function
\begin{eqnarray}
\Sigma (\omega ) &=&A^2\int_0^\infty dt\exp (i\omega t)\sum_{{\bf q}}\langle
S_{-{\bf q}}^z(t)S_{{\bf q}}^z+\frac 12S_{-{\bf q}}^{-}(t)S_{{\bf q}%
}^{+}\rangle , \\
\langle S_{-{\bf q}}^\alpha (t)S_{{\bf q}}^\beta \rangle &=&\int_{-\infty
}^\infty d\omega \exp (i\omega t){\cal J}_{{\bf q}}^{\alpha \beta }(\omega ),%
{\cal J}_{{\bf q}}^{\alpha \beta }(\omega )=-\frac 1\pi N(\omega )\text{Im}%
\langle \langle S_{{\bf q}}^\beta |S_{-{\bf q}}^\alpha \rangle \rangle
_\omega .
\end{eqnarray}
As discussed in Ref.\cite{IKJP91}, the spectral density ${\cal J}_{{\bf q}%
}^{\alpha \beta }(\omega )$ contains an almost singular contribution
\begin{equation}
\delta {\cal J}_{{\bf q}}^{\alpha \beta }(\omega )\propto \Delta _{{\bf q-Q}%
}\Delta _\omega
\end{equation}
where $\Delta _{{\bf q}}$ and $\Delta _\omega $ are delta-like functions
smeared at the scales $q\sim \xi ^{-1}$ and $\omega \sim \omega _\xi $ with
the characteristic spin-fluctuation energy $\omega _\xi \sim {\cal D}\xi
^{-2}$ (FM), $\omega _\xi \sim c\xi ^{-1}$ (AFM). To obtain the singular
term in $\Sigma (\omega )$ with the correct factor of $S^2$ we can introduce
a very small magnetic anisotropy (which does not violate time-reversal
symmetry), so that the whole singular contribution passes to $K_{{\bf q}%
}^{zz}(\omega )$ (cf. Ref.\cite{IKZ}). Then the term $iA^2S^2/\omega $
occurs at $\omega \gg $ $\omega _\xi $, and we obtain the expression
\begin{equation}
(I^{+},I^{-})^\omega =\frac i3I(I+1)\left( \frac 1{\omega -AS}+\frac 1{%
\omega +AS}\right)
\end{equation}
which describes precession of the nuclear spin with {\it both} frequencies $%
\pm AS$. We see that the resonance picture holds at $\omega _n\gg $ $\omega
_\xi $ only. In the opposite case the NMR line is smeared, but we can
calculate the quantity $1/T_1$ according to (\ref{tt1}).

Provided that $\omega _0\ll $ $\omega _\xi $ the quantity $\omega _\xi $
plays a role of the gap in the magnon spectrum. Therefore at $\omega _n\ll
\omega _\xi $ the cutoffs in the singular contributions to $1/T_1$ described
by (\ref{a12}), (\ref{a22}), (\ref{zz2d}) are determined by very small
inverse correlation length, so that they have very large values and possess
unusual temperature behavior.

In the $2D$ FM case the expression (\ref{f22}) is also applicable, but with
another expression for $\omega _0,\omega _0\rightarrow {\cal D}\xi ^{-2}$
which is exponentially small. We have 
\begin{equation}
\delta ^{(2)}(1/T_1)\propto I^2A^2T^2/{\cal D}^{5/2}\omega _\xi
^{1/2}\propto I^2\omega _n^2\xi T^2/{\cal D}^3.  \label{d21}
\end{equation}
In the isotropic $2D$ AFM case we obtain from (\ref{a12}) 
\begin{equation}
\delta ^{(1)}(1/T_1)\propto I^2A^2T/c\omega _\xi \propto I^2\omega _n^2\xi
T/c^2.  \label{d22}
\end{equation}
As follows from (\ref{a22}), (\ref{korr}), 
\begin{equation}
\delta ^{(2)}(1/T_1)\propto I^2A^2(T^2/c^4)\ln ^2\xi \simeq \text{const}(T).
\label{d23}
\end{equation}
Possibility to observe such dependences experimentally is of great interest.
Unfortunately, most experimental data for layered compunds deal with
copper-oxide systems which are on the bordeline of AFM instability. The
latter results in a specific temperature dependence of the spin
susceptibility and strong deviations from the Korringa law. This makes
separation of one- and two-magnon contributions impossible.

At $\omega _n\gg $ $\omega _\xi $ the cutoff in the singular contributions
to $1/T_1$ described by (\ref{a12}), (\ref{a22}), (\ref{zz2d}) is the NMR
frequency $\omega _n.$ However, such a cutoff is absent for the
corresponding terms in $1/T_2$ owing to the interaction $\widetilde{A}\sim
a|F^{(1)}|$. Provided that $\widetilde{A}\sim A$ we reproduce for these
terms the dependences (\ref{d21})-(\ref{d23}). However, as discussed in
Sect.2, for large $\xi $ the lineform turns out to be Gaussian owing to
strong Ruderman-Kittel interaction between nuclear spins,  the Moriya
formula (\ref{tt2}) is inapplicable, and one can estimate the linewidth from
(\ref{gauss}). In the 2D case we obtain $1/T_2\propto \xi .$ A more detailed
discussion of this situation and an application to copper-oxide systems is
given in Ref.\cite{thelen}.

We see that NMR investigations can be in principle used to obtain the
temperature dependence of the correlation length. When crossing the magnetic
ordering point in layered systems the NMR picture should not change radically%
$.$ Formally, as follows from (\ref{tm}), (\ref{korr}) at $T\sim T_M$ we
have $\ln \xi \simeq \ln |JS^2/\max \{\omega _0,|J^{\prime }|\}|,$ so that
the cutoffs are joined smoothly.

A similar situation can take place for other systems with suppressed LRO and
strong SRO, e.g., for frustrated 3D magnetic systems \cite{Kat} (where $\xi $
is also large, the lineform is Gaussian, and we obtain from (\ref{gauss}) $%
1/T_2\propto \xi ^{1/2}$). This may explain why the problem of detecting
long-range magnetic ordering is frustrated systems with small ordered
moments with the use of the NMR method so difficult. Indeed, the NMR data
for heavy-fermion systems are doubtful and contradict to results of other
experiments \cite{hf}.

\section{Conclusions}

In the present paper we have investigated in detail various mechanism of
nuclear magnetic relaxation in metallic ferro- and antiferromagnets in the
spin-wave temperature region. In the most cases the main contribution to $%
1/T_1$ is of Korringa type, but its physical origin is more complicated than
in paramagnetic metals. Formally, it results from the interaction of nuclear
magnetic moments with the {\it localized }electronic subsystem with taking
into account the ``Stoner'' (Landau) damping of spin waves via conduction
electrons. This contribution is greatly enhanced in comparison with the
standard Korringa term by inverse powers of exchange interaction ($s-d(f)$
parameter), especially in ferromagnets. In $3D$ antiferromagnets such a
contribution contains the logarithmic singularity which is cut at the gap in
the magnon spectrum (magnetic anisotropy energy) $\omega _0.$ Thus we can
conclude that the ``Korringa'' relaxation rate in magnetic metals should be
much larger than in paramagnetic ones where the relaxation is determined by
direct interaction with conduction electrons (such a term is also present in
the magnetically ordered state, but is much smaller than the contribution
discussed). In the $2D$ AFM case we have $1/T_1\propto \omega _0^{-1}$. In
the isotropic limit the singularity in $1/T_1$ is cut at very small inverse
correlation length, so that the one-magnon contribution becomes very large
and possesses unusual temperature behavior.

Besides that, we have calculated contributions from more complicated magnon
damping processes (electron-magnon scattering). For antiferromagnets and $2D$
ferromagnets they contain singular logarithmic or (in the $2D$ FM case)
power-law factors which are also cut at $\omega _0$. These contributions may
result in considerable deviations of the temperature dependence of $1/T_1$
from the linear Korringa law. In the $3D$ FM case this contribution is also
noticeable and probably can be separated when fitting experimental data. For
half-metallic ferromagnets, where the ``Stoner'' damping is absent, this
scattering should be the main nuclear relaxation mechanism (see the
discussion of experimental data in the review \cite{UFN}).

Provided that the ``longitudinal'' matrix elements of dipole-dipole
hyperfine interactions in (\ref{FK}) are not too small, the two-magnon
(``Raman'') relaxation processes may be also important in $1/T_1$,
especially for $2D$ ferromagnets.

The research described was supported in part by Grant No.99-02-16279 from
the Russian Basic Research Foundation.

\end{document}